\documentclass[english,twocolumn,superscriptaddress,longbibliography,floatfix]{revtex4-1}

\usepackage{graphicx}
\usepackage{amsmath,amsfonts}
\usepackage{hyperref}
\usepackage{tikz}
\usepackage{siunitx}
\usepackage{float}
\usepackage[title]{appendix}
\usepackage{xcolor}
\usepackage[english]{babel}
\usepackage{booktabs}
\usepackage{multirow}
\usepackage{textcomp}
\usepackage{lineno}

\definecolor{lime}{HTML}{A6CE39}
\DeclareRobustCommand{\orcidicon}{
  \begin{tikzpicture}
  \draw[lime, fill=lime] (0,0)
  circle [radius=0.16]
  node[white] {{\fontfamily{qag}\selectfont \tiny ID}};
  \draw[white, fill=white] (-0.0625,0.095)
  circle [radius=0.007];
  \end{tikzpicture}
  \hspace{-2mm}
}
\foreach \x in {A, ..., Z}{\expandafter\xdef\csname orcid\x\endcsname{\noexpand\href{https://orcid.org/\csname
      orcidauthor\x\endcsname}{\noexpand\orcidicon}} }

\begin{document}

\title{Design and optimization of the NSLS-II lattice with complex bend replacement}

\author{Minghao Song\orcidA{}}\email{msong1@bnl.gov}
\author{Yoshiteru Hidaka\orcidB{}}
\author{Bernard Kosciuk} 
\author{Patrick N'Gotta\orcidD{}} 
\author{\protect\linebreak Sushil Sharma\orcidF{}}
\author{Timur Shaftan}  
\author{Guimei Wang} 
\affiliation{Brookhaven National Laboratory, Upton, New York 11973, USA}

\begin{abstract}

The demand for brighter photon beams has driven the development of diffraction-limited storage rings with ultralow electron beam emittance. A complex bend (CB) lattice was recently proposed as an approach for the future NSLS-II upgrade. As a step toward this upgrade, NSLS-II plans to replace two existing electromagnetic dipoles with permanent-magnet CBs, providing the first experimental demonstration of the CB concept in a storage ring. This paper presents the linear optics design, nonlinear optimization, and machine error evaluation of the modified NSLS-II lattice. The optimized lattice provides sufficient dynamic and momentum apertures for off-axis injection and routine operation and can be corrected with the existing NSLS-II correction scheme. Scraper measurements near the planned CB installation location validate the beam aperture requirements used in the CB design.

\end{abstract}

\maketitle

\section{Introduction}

Synchrotron light sources are moving toward diffraction-limited storage rings to provide brighter and more coherent photon beams. Electron beam emittance, which strongly affects the photon beam brightness, scales approximately with the inverse cube of the number of bending magnets in a storage ring~\cite{Murphy_E2N3}. Increasing the number of bending magnets is therefore an effective way to reduce the natural emittance from the nanometer to the picometer range. This principle has led to the broad use of multibend-achromat (MBA) lattices~\cite{Einfeld_DIFL} in next-generation storage-ring light sources, including MAX IV~\cite{Leeman_MAXIV}, ESRF-EBS~\cite{Raimondi_ESRF-EBS}, APS-U~\cite{Fornek_APS-U}, ALS-U~\cite{Steier_ALS-U}, Elettra II~\cite{ElettraII}, and HEPS~\cite{Jiao_HEPS}.

The National Synchrotron Light Source II (NSLS-II) is a third-generation synchrotron light source that currently operates 30 beamlines covering a broad spectral range from infrared to hard x rays. Its accelerator includes a 3~GeV electron storage ring with a horizontal emittance of approximately 1~nm$\cdot$rad when three damping wigglers are in operation~\cite{NSLS-II_design_report}. To maintain its scientific capabilities and remain competitive with next-generation light sources, NSLS-II is studying an upgrade of its existing double-bend-achromat (DBA) lattice~\cite{chasman_DBA}.

An MBA lattice is a viable option for the future NSLS-II upgrade. However, pushing the emittance closer to the diffraction limit within the existing tunnel requires many bending magnets and leaves less space for insertion devices (IDs) and other accelerator components. For example, one proposed MAX IV upgrade lattice uses a 19-bend achromat and reaches a natural emittance of 16~pm$\cdot$rad within the existing 500-m ring tunnel~\cite{Tavares_MAXIV_future}. This is a factor of 20 below the emittance of the present MAX IV storage ring, but the large number of magnets leaves little space for IDs and other equipment.

This example illustrates a limitation of the MBA approach when it is applied within an existing tunnel. Therefore, it motivates the development of new magnet and lattice concepts. One such concept is the complex bend (CB), first proposed for the future NSLS-II upgrade~\cite{Timur_2018_techreport}. The CB concept has evolved through several stages. The first design used a series of short dipoles interleaved with strong quadrupoles~\cite{Guimei_CBI}. A later design shifted the quadrupoles transversely to generate dipole fields~\cite{Guimei_CBII}. The current design uses permanent-magnet dipole-quadrupole combined-function magnets of the Halbach or hybrid type~\cite{Patrick_ESRF_PMQ,Sushil_retreat_slides,Patrick_napac2025,Patrick_talk}. Based on this design, low-emittance storage ring lattices have been developed for the NSLS-II upgrade at beam energies of 3~GeV~\cite{Minghao_CBA} and 4~GeV~\cite{song:ipac2024-thpc82}.

To demonstrate the CB concept, a CB prototype was previously built and tested in the NSLS-II linac beamline~\cite{Guimei_CB_prototype}. The beam energy was scaled from 100 to 200~MeV while maintaining strong focusing. The prototype used a symmetric 16-wedge Halbach design and reached a gradient of 140~T/m in compact permanent-magnet quadrupoles.

\begin{figure}[!htb]
   \centering
   \includegraphics*[width=0.9\columnwidth]{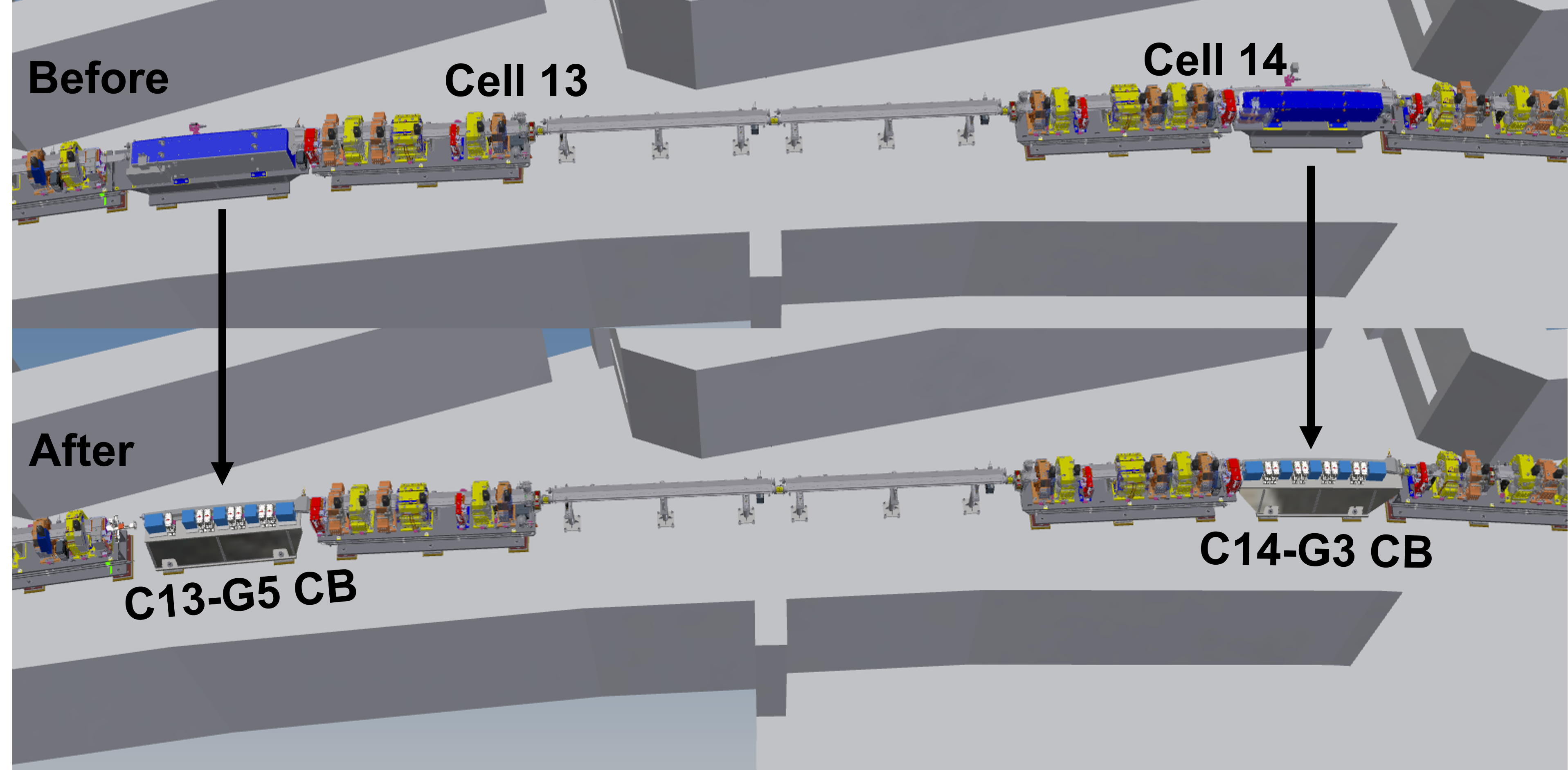}
   \caption{Schematic layout of the dipole replacement with complex bends in Cells 13 and 14 of the NSLS-II storage ring.}
   \label{dipole_CB_swap}
\end{figure}

To further validate the CB approach as a reliable, high-performance solution for the NSLS-II upgrade, a storage ring installation is the next step in testing CB under realistic operating conditions, including long-term exposure to temperature changes and radiation. A new project has therefore been started to replace two existing electromagnetic dipoles in the NSLS-II storage ring with permanent-magnet CBs, as shown in Fig.~\ref{dipole_CB_swap}. The project will provide a long-term demonstration of the CB concept in an operating 3~GeV storage ring and will allow beam dynamics studies using the existing NSLS-II beam position monitors (BPMs).

The two selected dipoles form a locally mirror-symmetric pair around a high-beta straight. However, replacing only these two dipoles breaks the periodic symmetry of the full NSLS-II lattice. Simulation and experimental studies of an asymmetric NSLS-II lattice are reported in Ref.~\cite{song:napac2025_AsymmetircalLattice}. To minimize the impact on routine operation, the modified lattice must preserve high injection efficiency and adequate beam lifetime.

In this work, we develop an NSLS-II lattice with two CB replacements and optimize its nonlinear beam dynamics. We then evaluate the designed lattice with realistic alignment, field, and multipole errors. The study also includes an experimental measurement of the beam aperture near the planned CB location. Together, these results support the planned CB installation and provide information for a future CB-based NSLS-II upgrade.

The CB installation project includes lattice design, magnet development, fabrication, installation, and beam commissioning. This paper focuses on the lattice design and beam dynamics studies. The other project activities are beyond the scope of this paper and will be reported separately.

The paper is organized as follows. Section~\ref{sec:Project objectives and design constraints} describes the project objectives and design constraints. Section~\ref{sec:Lattice design} presents the linear optics design and nonlinear optimization of the lattice with CBs. Section~\ref{sec:Performance with machine imperfections} evaluates the lattice with realistic machine errors and discusses the correction capability and magnetic cross-talk. Section~\ref{sec:Experimental validation of the CB beam aperture} presents the experimental measurements of the beam apertures. Section~\ref{sec:Conclusion} concludes the paper.

\section{Project objectives and design constraints}
\label{sec:Project objectives and design constraints}

The CB installation project includes the following objectives: (i) Develop and demonstrate the accelerator design of CBs and their integration into the NSLS-II ring lattice. (ii) Study the major impact of mechanical and magnetic constraints on the ring lattice through beam studies, using scrapers and 8-pole magnets to simulate the changes imposed by CBs in the ring. (iii) Build and characterize two CBs and verify that they meet the required specifications using mechanical and magnetic measurement benches. (iv) Install the CBs in the tunnel and prepare the ring for operation in a single shutdown. (v) Characterize the performance of the CBs and the modified ring lattice during commissioning and operation.

Meanwhile, the project must preserve the existing NSLS-II infrastructure as much as possible. This requirement leads to the following lattice design constraints: (i) Only the two dipole girders are replaced, and the beam coordinates at the entrance and exit of each replacement section must remain unchanged so that the equipment on neighboring girders does not need to be modified. (ii) Local CB optics must be matched to the ring while the gradients of nearby quadrupoles remain below 22~T/m. (iii) The modified lattice must provide sufficient dynamic aperture (DA) and momentum aperture (MA) for off-axis injection and adequate beam lifetime. (iv) Changes to the sextupole power supply system must be limited by using three local power supply pairs for the six harmonic sextupoles near the CBs, independent of the shared pentant circuit. (v) The lattice must remain tolerant to machine imperfections.

The achievable CB fields and gradients are limited by the required beam aperture and by permanent-magnet technology. The minimum aperture is obtained from the injection DA requirement and is scaled from the injection septum to the CB location. This gives a magnet bore radius of 11.5~mm. For the hybrid permanent-magnet design considered here, the maximum absolute pole-tip field ($|{B}_\text{max}|$) is approximately 1.4~T. It is estimated from
\begin{equation}
    |B_{\text{max}}| = |B| + |G|r,
\end{equation}
where $B$ is the dipole field, $G$ is the quadrupole gradient, and $r$ is the magnet bore radius. This limit constrains the available dipole field and quadrupole gradient of the CB elements.

The spacing between neighboring permanent magnets is also important for magnetic design and mechanical integration. A minimum separation of 3~cm is reserved to limit magnetic cross-talk and to provide space for the magnet structures. The magnetic cross-talk is discussed further in Sec.~\ref{sec:cross_talk}.

\section{Lattice design}
\label{sec:Lattice design}

\subsection{Linear optics design}

The modified lattice was developed from the operational NSLS-II 17-ID lattice. The dipoles on Girder 5 of Cell 13 and Girder 3 of Cell 14 are replaced by CBs. These locations are adjacent to the same high-beta straight and form a locally mirror-symmetric pair, which helps reduce the local optics perturbation.

Each original NSLS-II dipole is 2.62~m long and provides a bending angle of $6^\circ$. Because of the project cost constraints, each replacement CB consists of three types of permanent magnets rather than only combined-function magnets: focusing combined-function magnets (CB-BQF), defocusing combined-function magnets (CB-BQD), and pure dipoles (CB-B). Together, these elements provide the required $6^\circ$ bending angle while preserving the original floor coordinates.

\begin{figure}[!htb]
   \centering
   \includegraphics[width=0.8\columnwidth]{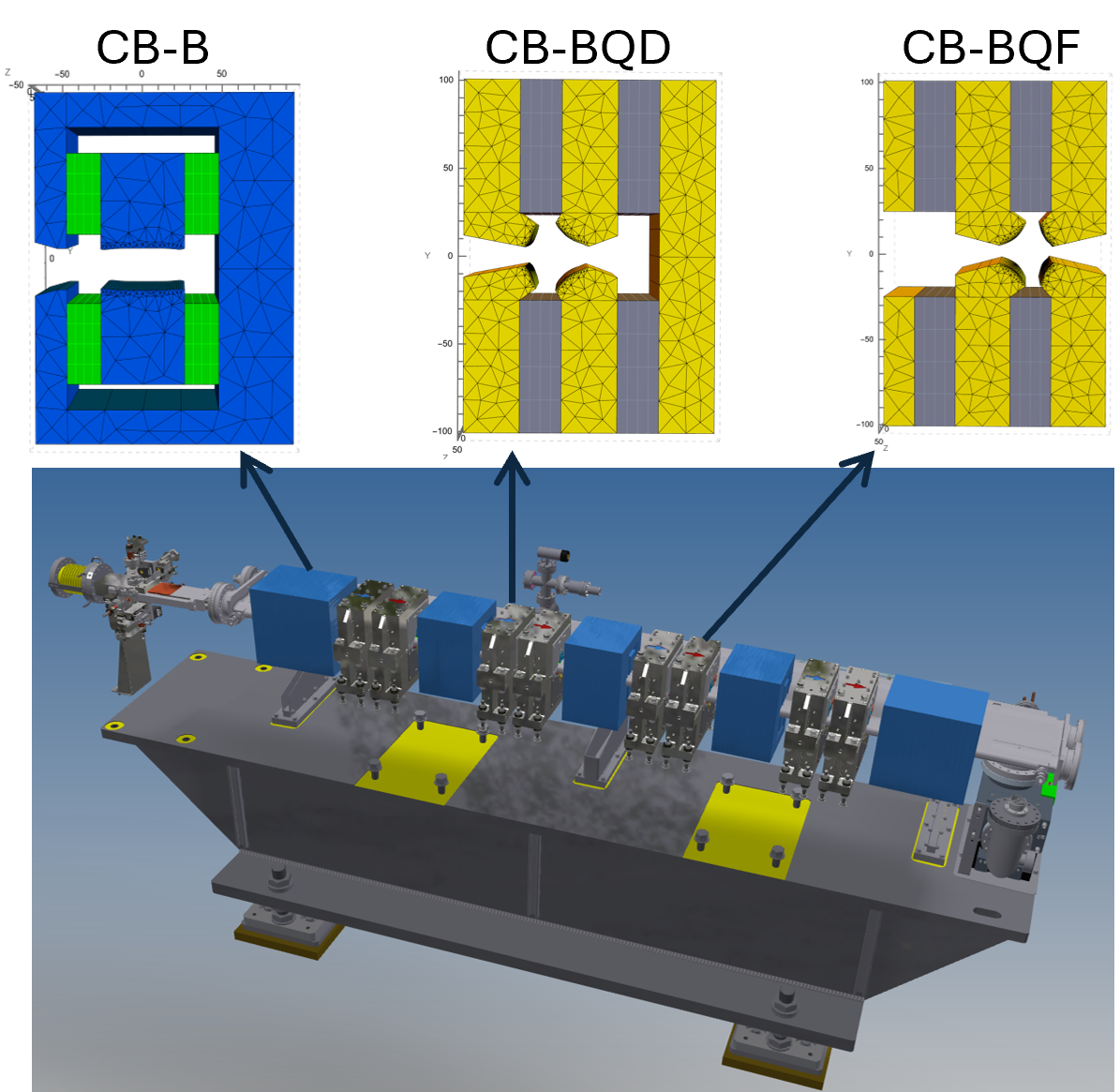}
   \caption{Schematic layout of one CB magnet. The CB-BQF and CB-BQD combined-function magnets are each 10~cm long. The CB-B pure dipoles are 30~cm long at the two ends and 20~cm long elsewhere.}
   \label{complex_bend_array}
\end{figure}

The CB layout is shown in Fig.~\ref{complex_bend_array}. The adjacent elements CB-BQF and CB-BQD are separated by 3~cm, consistent with the spacing considered for the future NSLS-II upgrade. A CB-B element is placed next to each CB-BQD and CB-BQF pair, with a 5-cm gap between the pure dipole and each neighboring combined-function magnet. To reduce the required dipole field, the CB-B elements at the two ends are 30~cm long, while the interior CB-B elements are 20~cm long.

The total length of the CB is 2.52~m, leaving 10~cm within the original 2.62-m dipole space for coordinate matching. After matching, the beam coordinates at the entrance and exit differ from those of the original dipole by less than 1.6~\textmu m.

The CB replacement changes the local optics. The strengths of the CB-BQF and CB-BQD elements and the nearby electromagnetic quadrupoles are therefore varied to match the optics while keeping all gradients within their limits. Small additional adjustments restore the betatron tunes and chromaticities to the values of the nominal 17-ID lattice.

Table~\ref{CB specifications} lists the CB magnet parameters obtained from the coordinate and optics matching. The CB magnet requires stronger fields and gradients than the existing dipoles and quadrupoles, but their calculated pole-tip fields remain below the 1.4-T design limit and are comparable to those considered for the future NSLS-II upgrade.

\begin{table}[!hbt]
   \centering
   \caption{CB magnet specifications.}
   \begin{tabular}{lcccc}
       \hline
       Element & $L$ (cm) & $B$ (T) & $G$ (T/m) & $\vert$$B_{\text{max}}$$\vert$ (T) \\
       \hline
       CB-B   & 20/30 & 0.52 & --     & --   \\
       CB-BQD & 10    & 0.66 & -53.86 & 1.28 \\
       CB-BQF & 10    & 0.38 & 60.41  & 1.07 \\
       \hline
   \end{tabular}
   \label{CB specifications}
\end{table}

\begin{figure}[!htb]
   \centering
   \includegraphics[width=1\columnwidth]{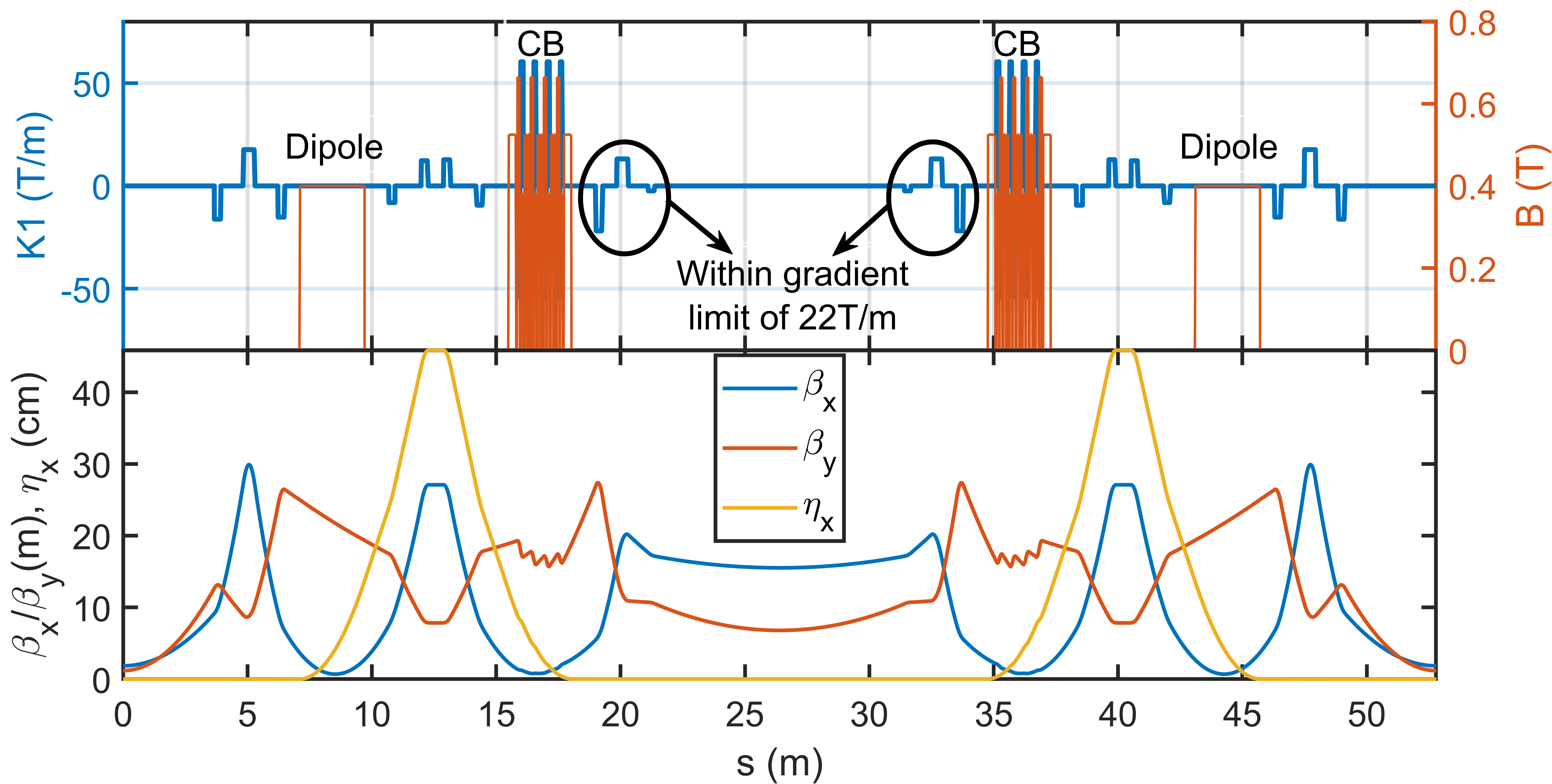}
   \caption{Magnetic field of CBs and the corresponding Twiss functions in Cells 13 and 14.}
   \label{lattice with CB}
\end{figure}

Figure~\ref{lattice with CB} shows the CB field profile and the corresponding Twiss functions in Cells 13 and 14. The beta functions oscillate through the CB because of the alternating focusing and defocusing gradients. The main ring parameters of the 17-ID lattice with CBs are listed in Table~\ref{Lattice parameters}. Because only two dipoles are replaced, the CBs have only a small effect on the overall ring emittance.

\begin{table}[!hbt]
   \centering
   \caption{Main parameters of the NSLS-II 17-ID lattice with two CB replacements.}
   \begin{tabular}{ll}
       \hline
       Parameter & Value \\
       \hline
       Beam energy (GeV) & 3 \\
       Circumference (m) & 791.958 \\
       Number of cells & 30 \\
       Natural emittance (pm$\cdot$rad) & 752.1 \\
       Damping partitions $(J_x,J_y,J_\delta)$ & (1.01, 1.00, 1.99) \\
       Ring tunes $(\nu_x,\nu_y)$ & (33.222, 16.263) \\
       Natural chromaticities $(\xi_x,\xi_y)$ & (-97.780, -40.741) \\
       Corrected chromaticities $(\xi_x,\xi_y)$ & (+2.060, +2.376) \\
       Momentum compaction & $3.62\times10^{-4}$ \\
       Energy spread (\%) & 0.080 \\
       Energy loss per turn (MeV) & 0.835 \\
       \hline
   \end{tabular}
   \label{Lattice parameters}
\end{table}

\subsection{Nonlinear dynamics optimization}

\begin{figure}[!htbp]
   \centering
   \includegraphics[width=0.8\columnwidth]{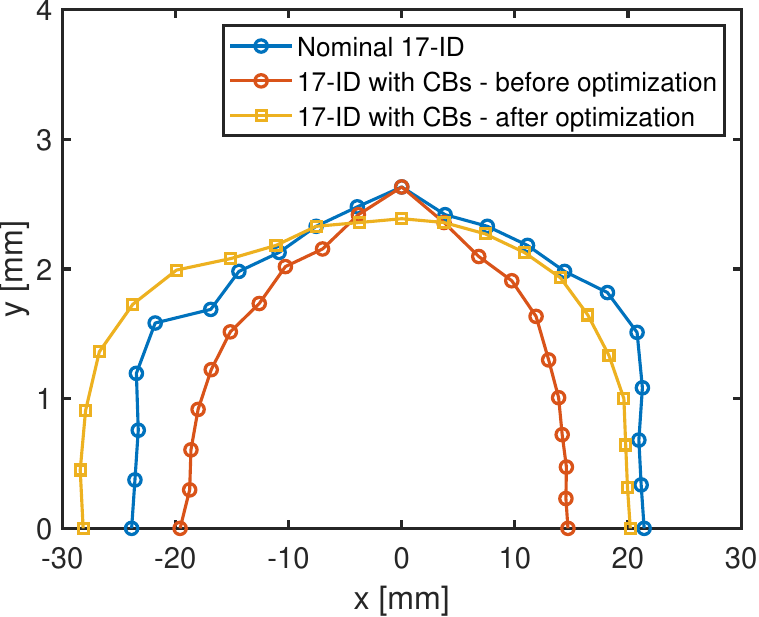}
   \caption{Comparison of the DA for the nominal 17-ID lattice and the 17-ID lattice with CBs, before and after optimization at the injection point.}
   \label{cmp_DA}
\end{figure}

The CB replacements change the linear optics and reduce the periodic symmetry of the 17-ID lattice. As a result, the performance of nonlinear beam dynamics is degraded before optimization, as shown in Fig.~\ref{cmp_DA}. The DA and MA are therefore optimized to recover the acceptance required for injection and routine operation.

Only harmonic sextupoles are used in this optimization, so that the first-order chromaticities remain unchanged. The NSLS-II ring has six harmonic-sextupole families: SL1, SL2, and SL3 in the low-beta straights, and SH1, SH3, and SH4 in the high-beta straights. Three pairs of sextupoles near the CBs are powered by independent local supplies and form three local optimization knobs. The remaining SH1, SH3, and SH4 sextupoles form three additional knobs, and the SL1, SL2, and SL3 sextupoles form another three. The optimization therefore uses nine sextupole knobs in total.

During optimization, each harmonic sextupole strength is constrained to the current operational limit of 40~T/m$^2$. The DA and MA are optimized simultaneously, where the DA is evaluated as the stable area in the transverse x-y plane and the MA is evaluated as the stable momentum deviation at the injection point. The vertical DA is limited by the ID gap. The final solution is selected based on its overall DA and MA performance. As shown in Fig.~\ref{cmp_DA}, the optimized sextupole settings restore the DA of the 17-ID lattice with CBs to a level comparable to that of the nominal 17-ID lattice.

\begin{figure}[!htbp]
   \centering
   \includegraphics[width=0.8\columnwidth]{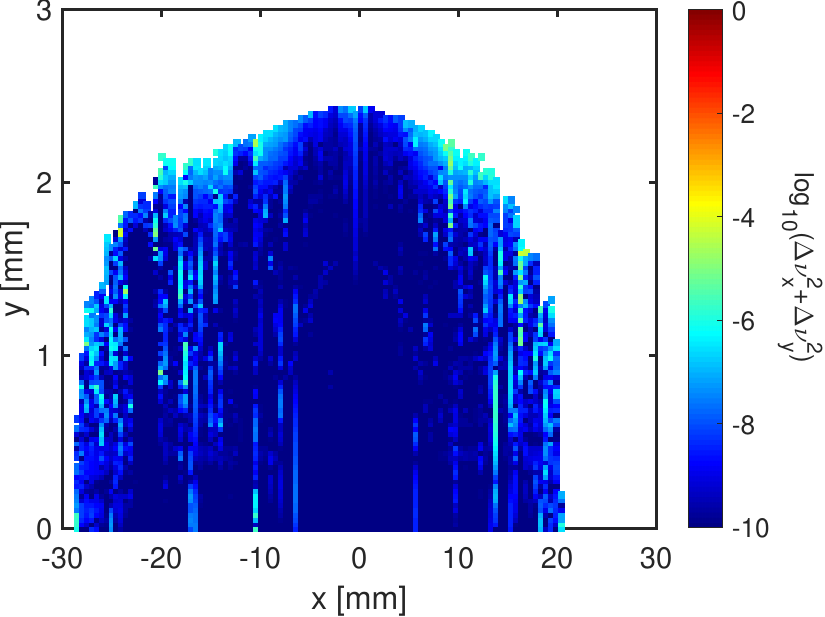}
   \includegraphics[width=0.8\columnwidth]{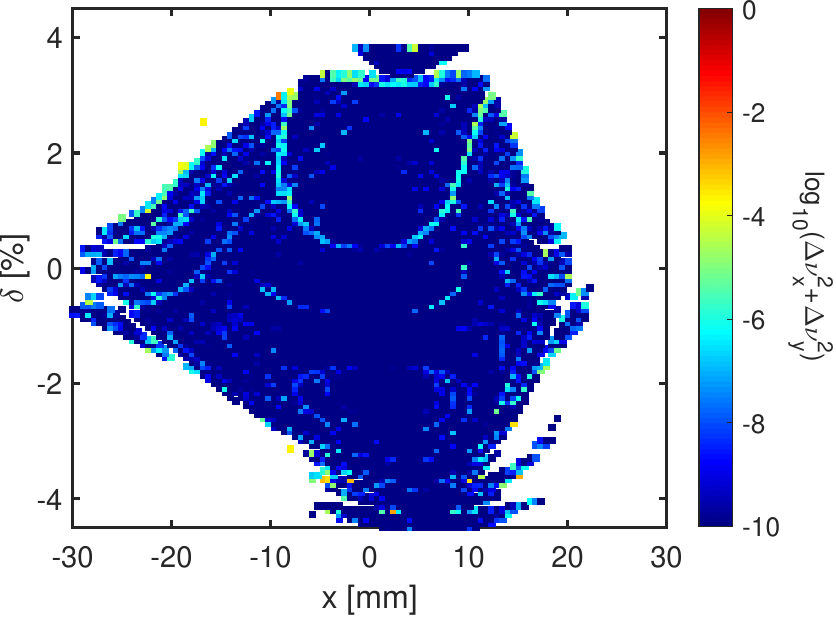}
   \caption{Frequency maps at the injection point for the optimized 17-ID lattice with CBs: on-momentum DA in the x-y plane (upper) and off-momentum acceptance in the x-$\delta$ plane (lower). The color scale denotes the tune diffusion rate $\log_{10}\sqrt{\Delta\nu_x^2+\Delta\nu_y^2}$~\cite{laskar1999introduction}.}
   \label{fmap}
\end{figure}

In accelerator physics, the on-momentum DA is commonly defined by its two-dimensional projection in the transverse x-y plane. A sufficiently large DA is required to accommodate the injected beam in the presence of finite beam size, energy spread, magnet misalignment, and other lattice imperfections. The two-dimensional projection in the horizontal-momentum (x-$\delta$) plane provides complementary information on the acceptance of off-momentum particles. 
Figure~\ref{fmap} shows both projections for the optimized lattice. The color scale denotes the tune diffusion rate~\cite{laskar1999introduction}. The optimized lattice provides sufficient on- and off-momentum DA.

\section{Performance with machine imperfections}
\label{sec:Performance with machine imperfections}

\subsection{Error model and correction scheme}

Magnet misalignments and field errors distort the orbit and optics, shift the tunes, increase coupling, and reduce the dynamic and momentum apertures. Therefore, the performance of the designed 17-ID lattice with CBs must be evaluated under realistic machine imperfections using numerical tracking simulations. All tracking studies in this section are performed with \textsc{elegant}~\cite{Borland_elegant}.

\begin{table*}[!t]
\centering
\caption{Machine imperfections used in the performance evaluation. The girder, electromagnetic quadrupole (QUAD and HIQUAD), and sextupole (SEXT and HISEXT) errors follow the existing NSLS-II specifications. The 1$\sigma$ cutoff is the cutoff used in the existing NSLS-II error model. The CB error values are based on current fabrication and alignment capabilities. FSE denotes fractional strength error.}
\label{tab_error_params}
\begin{tabular}{llccc}
    \hline
    Component & Error parameter & RMS & Unit & Cutoff ($\sigma$) \\
    \hline
    \multirow{6}{*}{CB-BQD/CB-BQF}
      & Offset $x/y$       & \num{30}   & \textmu m & \num{1.0} \\
      & Roll               & \num{0.5}  & mrad      & \num{1.0} \\
      & Dipole FSE         & \num{0.1}\% & --        & \num{1.0} \\
      & Quadrupole FSE     & \num{1}\%   & --        & \num{1.0} \\
      & $K_2$ (sextupole)  & \num{0.7}  & m$^{-3}$  & \num{1.0} \\
      & $K_3$ (octupole)   & \num{150}  & m$^{-4}$  & \num{1.0} \\
    \hline
    \multirow{5}{*}{CB-B}
      & Offset $x/y$      & \num{30}   & \textmu m & \num{1.0} \\
      & Roll              & \num{0.5}  & mrad      & \num{1.0} \\
      & Dipole FSE        & \num{0.1}\% & --        & \num{1.0} \\
      & $K_2$ (sextupole) & \num{0.7}  & m$^{-3}$  & \num{1.0} \\
      & $K_3$ (octupole)  & \num{150}  & m$^{-4}$  & \num{1.0} \\
    \hline
    \multirow{3}{*}{QUAD/HIQUAD}
      & Offset $x/y$ & \num{30}     & \textmu m & \num{1.0} \\
      & Roll         & \num{0.2}    & mrad      & \num{1.0} \\
      & FSE          & \num{0.025}\% & --        & \num{1.0} \\
    \hline
    \multirow{3}{*}{SEXT/HISEXT}
      & Offset $x/y$ & \num{30}     & \textmu m & \num{1.0} \\
      & Roll         & \num{0.2}    & mrad      & \num{1.0} \\
      & FSE          & \num{0.05}\% & --        & \num{1.0} \\
    \hline
    \multirow{2}{*}{Girders}
      & Upstream/downstream offset $x/y$ & \num{100} & \textmu m & \num{1.0} \\
      & Roll                               & \num{0.5} & mrad      & \num{1.0} \\
    \hline
\end{tabular}
\end{table*}

The errors used to generate the random lattices are summarized in Table~\ref{tab_error_params}. For all existing NSLS-II components, the error specifications are identical to those used for the operational 17-ID lattice, including the rms values and the 1$\sigma$ cutoff. This allows the performance of the lattice with CBs to be compared directly with that of the nominal 17-ID lattice. The CB elements use separate alignment, field, and multipole error specifications based on current fabrication and alignment capabilities.

Systematic and random multipole errors up to the 14th order are included for the electromagnetic quadrupoles and sextupoles. QUAD and SEXT elements are located in dispersive regions, while HIQUAD and HISEXT elements are located in non-dispersive regions. The CB error model includes transverse offsets, roll, dipole and quadrupole FSEs, and sextupole and octupole components. In addition, the effects of the insertion devices are also included by incorporating ID kick maps into the lattice.

The performance study uses 100 random error seeds. For each seed, orbit correction is applied first, followed by linear optics and coupling correction. The corrected lattices are then used to evaluate the DA, local momentum aperture (LMA), and Touschek lifetime.

\subsection{Orbit correction}

Orbit correction is performed using an orbit response matrix calculated from the designed 17-ID lattice with CBs. This design response matrix is then applied to all 100 random error seeds. The existing storage ring has 360 correctors and 360 BPMs, with 180 correctors and 180 BPMs in each transverse plane. The response matrix is used to determine the dipole corrector strengths that minimize the closed orbit error.

\begin{figure}[!htbp]
   \centering
   \includegraphics[width=0.8\columnwidth]{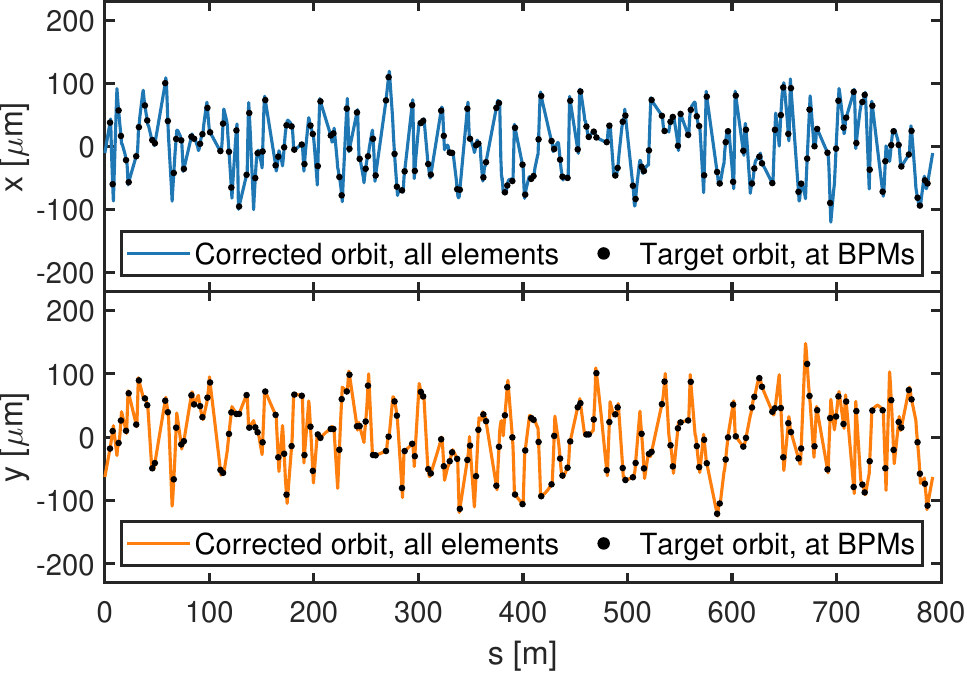}
   \caption{Corrected closed orbit for one random error seed. The upper and lower subplots show the horizontal and vertical planes, respectively. Black dots mark the target BPM positions, and the colored curves show the orbit at all lattice elements.}
   \label{orbit_one_instance}
\end{figure}

Figure~\ref{orbit_one_instance} shows the corrected orbit for one representative random error seed. 
BPM gain and roll errors are not included in the simulated corrections because the existing BPMs will be used and their calibrations are already known. These errors can be corrected. BPM offsets are also excluded because the orbit correction targets will be set to the known beam-based alignment (BBA) values. BPM noise is not included in this analysis for simplicity. 
The black dots show the correction target values at the BPM locations, while the colored curves show the orbit at all lattice elements. After orbit correction, the BPM readings are restored close to their target values. 

\begin{figure}[!htbp]
   \centering
   \includegraphics[width=0.8\columnwidth]{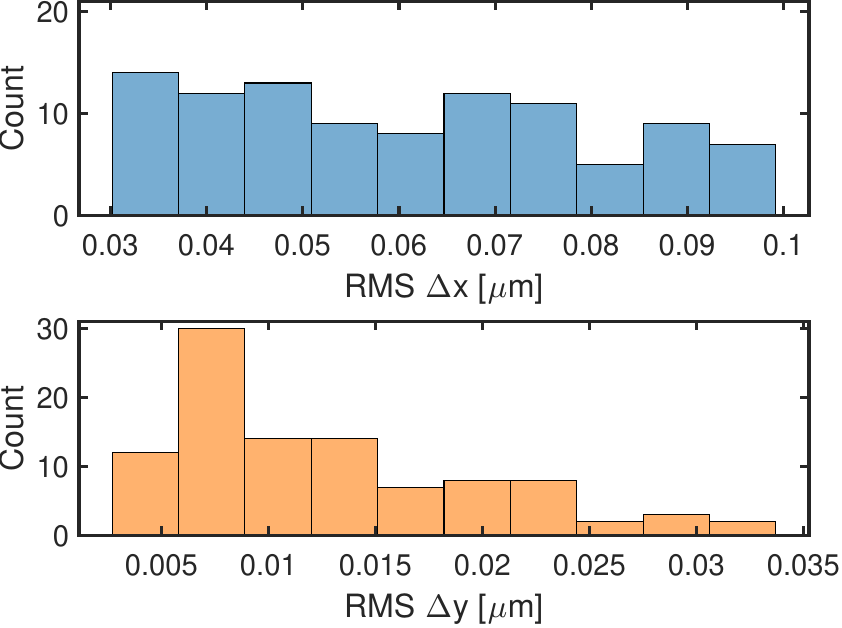}
   \includegraphics[width=0.8\columnwidth]{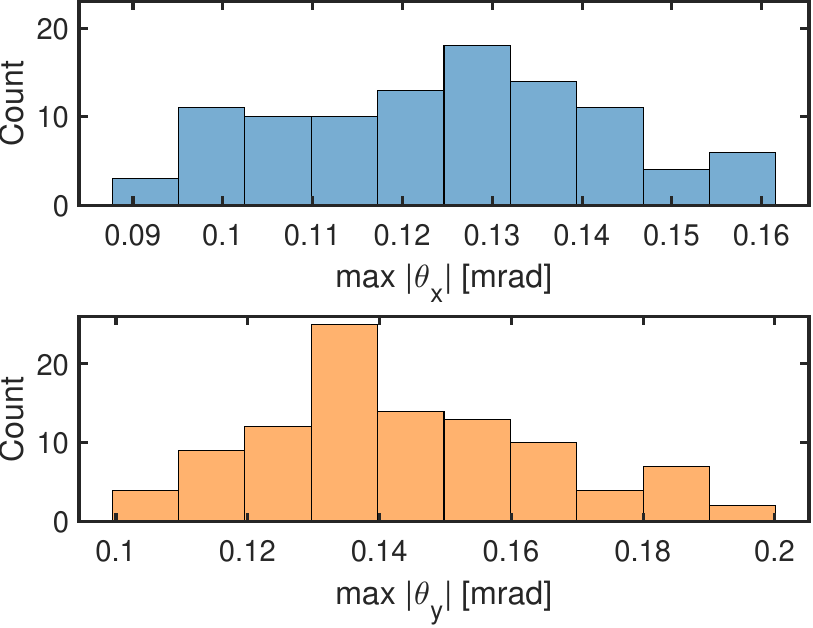}
   \caption{Orbit correction results for 100 random error seeds. From top to bottom: horizontal and vertical rms residual orbit, followed by the maximum horizontal and vertical corrector kicks.}
   \label{orbit_corrector_energy_after_correction}
\end{figure}

Figure~\ref{orbit_corrector_energy_after_correction} summarizes the orbit correction results for all 100 seeds. The rms residual orbit is below 0.1~\textmu m horizontally and 0.035~\textmu m vertically. The largest required corrector kick is below 0.2~mrad in both planes and remains within the practical operating range.

\subsection{Linear optics and coupling correction}

After orbit correction, the linear optics and transverse coupling are corrected using the design response matrix constructed from the primary and secondary spectral peaks of the simulated turn-by-turn data, together with the tune and dispersion.
All 300 normal quadrupoles and all 30 skew quadrupoles are included in the correction. The normal quadrupoles mainly correct the beta functions, tunes, and horizontal dispersion, while the skew quadrupoles correct the coupling and vertical dispersion.

\begin{figure}[!htbp]
   \centering
   \includegraphics[width=0.8\columnwidth]{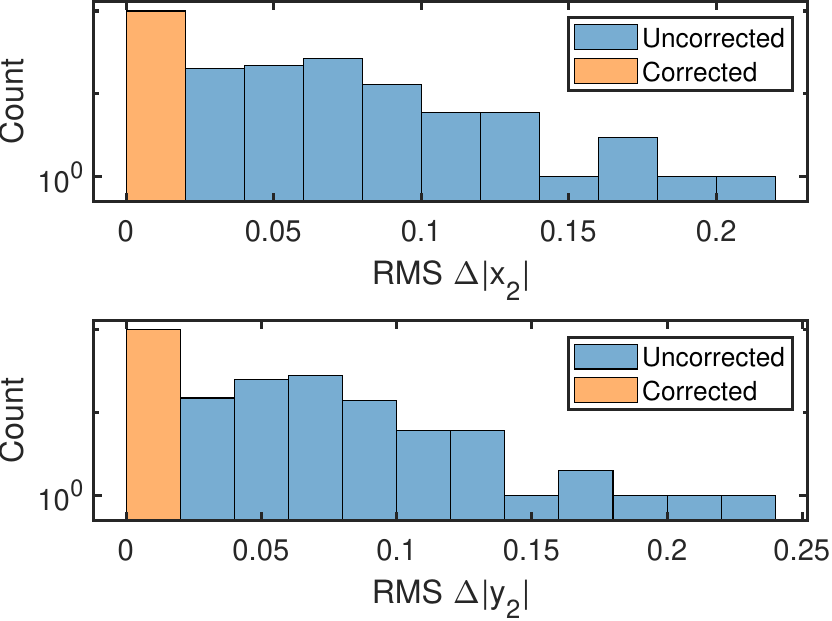}
   \caption{Distributions of the horizontal (upper) and vertical (lower) secondary spectral peaks before and after coupling correction for 100 random error seeds.}
   \label{rms_secondary_spectral_peaks}
\end{figure}

Transverse coupling is evaluated using the secondary spectral peaks obtained from the simulated turn-by-turn data. These peaks arise from coupling between the horizontal and vertical betatron motion and therefore provide a direct measure of transverse coupling. Figure~\ref{rms_secondary_spectral_peaks} shows their distributions before and after correction for the 100 random error seeds. The significant reduction of the secondary peaks demonstrates the effectiveness of the coupling correction.

\begin{figure}[!htbp]
   \centering
   \includegraphics[width=0.8\columnwidth]{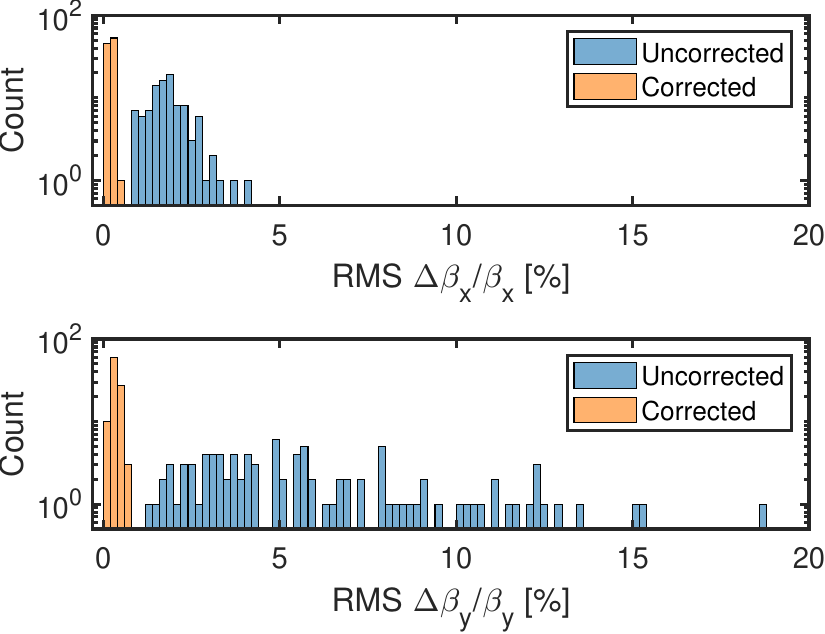}
   \includegraphics[width=0.8\columnwidth]{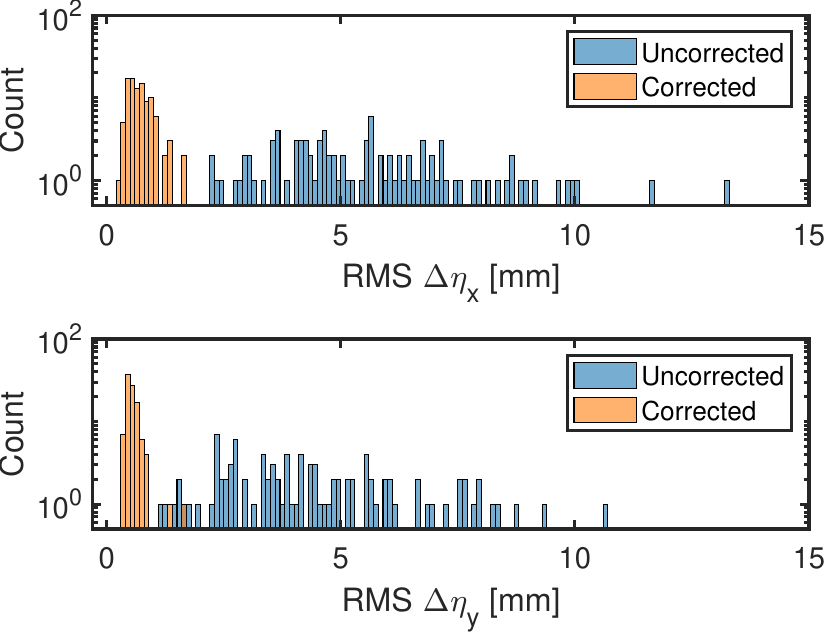}
   \caption{Linear optics correction results for 100 random error seeds. From top to bottom: horizontal and vertical rms beta beating, followed by horizontal and vertical rms residual dispersion. Results before and after correction are compared.}
   \label{beta_beat_residual_dispersion}
\end{figure}

\begin{figure}[!htbp]
   \centering
   \includegraphics[width=0.8\columnwidth]{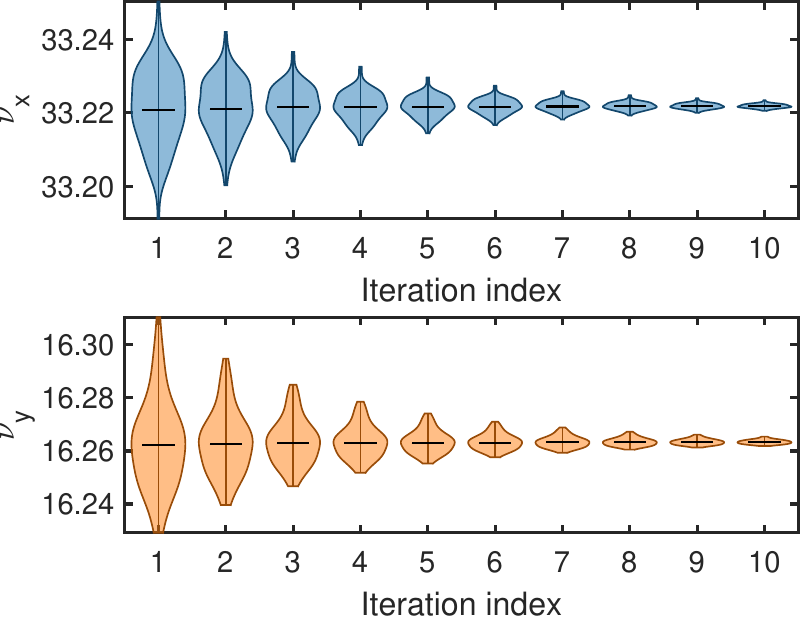}
   \caption{Horizontal (upper) and vertical (lower) betatron tunes over 100 random error seeds during 10 correction iterations.}
   \label{tune}
\end{figure}

Figure~\ref{beta_beat_residual_dispersion} summarizes the optics correction results for the 100 random error seeds. After correction, the rms beta beating is below 1\%, and the rms residual dispersion is below 2~mm in both planes. Consequently, the increase in beam size due to the residual dispersion is negligible. As shown in Fig.~\ref{tune}, the betatron tunes are also restored to their designed values. 

\begin{figure}[!htbp]
   \centering
   \includegraphics[width=0.8\columnwidth]{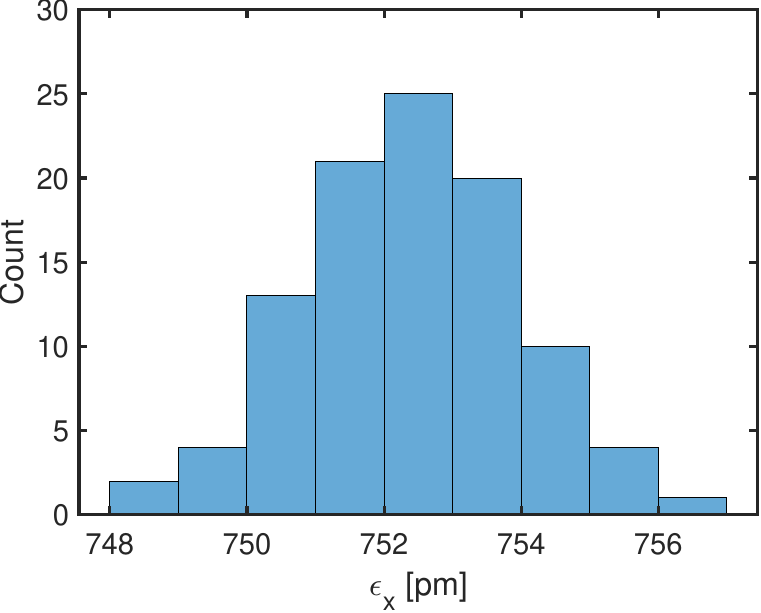}
   \caption{Distribution of the corrected horizontal emittance over 100 random error seeds.}
   \label{emittance}
\end{figure}

Figure~\ref{emittance} shows that the corrected horizontal emittance remains close to the design value listed in Table~\ref{Lattice parameters}. This result indicates that the corrected lattice preserves the designed beam emittance in the presence of realistic machine imperfections.

\subsection{Nonlinear beam dynamics after correction}

Following orbit, optics, and coupling corrections, the nonlinear beam dynamics of the corrected 17-ID lattice with CBs are evaluated using the same 100 random error seeds. To evaluate the impact of replacing two dipoles with CBs, the dynamic aperture (DA) and local momentum aperture (LMA) are compared with those of the corrected nominal 17-ID lattice.

\begin{figure}[!htb]
   \centering
   \includegraphics[width=0.8\columnwidth]{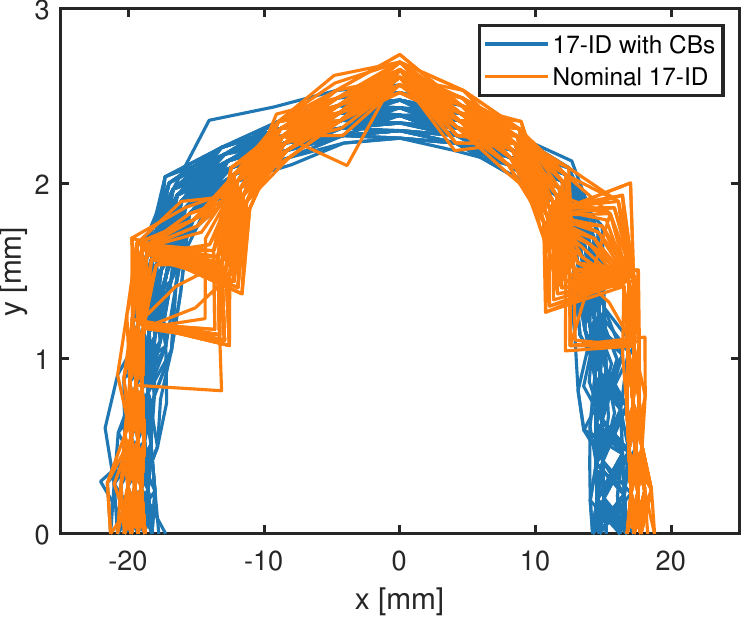}
   \includegraphics[width=0.8\columnwidth]{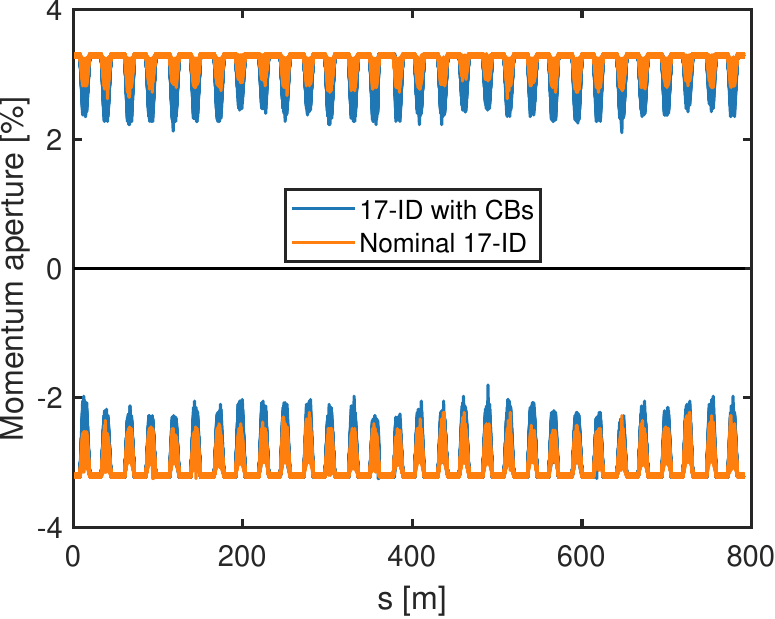}
   \caption{Comparison of the on-momentum dynamic aperture at the injection point (upper) and local momentum aperture around the ring (lower) for the corrected 17-ID lattices with CBs and the corrected nominal 17-ID lattices over 100 random error seeds.}
   \label{DA_LMA}
\end{figure}

Figure~\ref{DA_LMA} compares the on-momentum DA at the injection point and the LMA around the ring for the corrected 17-ID lattice with CBs and the corrected nominal 17-ID lattice over the 100 error seeds.
As shown in Fig.~\ref{DA_LMA}, the corrected lattice with CBs provides a DA comparable to that of the corrected nominal 17-ID lattice. The LMA is calculated by particle tracking with an rf voltage of 3.6~MV and a harmonic number of 1320. Although the lattices with CBs have a slightly smaller LMA than the nominal lattices, the momentum acceptance is sufficient for routine operation.

\begin{figure}[!htbp]
   \centering
   \includegraphics[width=0.8\columnwidth]{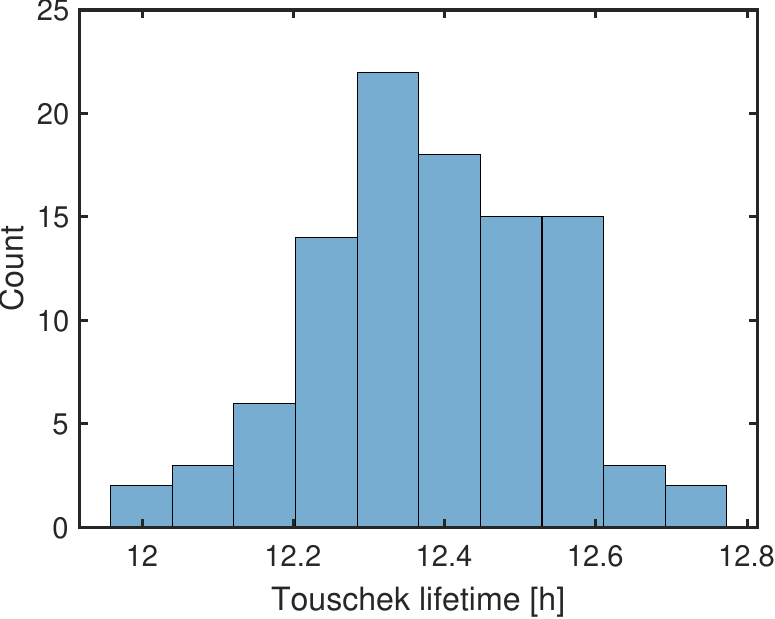}
   \caption{Distribution of the calculated Touschek lifetime for the corrected 17-ID lattices with CBs over 100 random error seeds. The average lifetime is approximately 12.4~h.}
   \label{Touschek_lifetime}
\end{figure}

The Touschek lifetime is then evaluated only for the corrected lattices with CBs using the same 100 random error seeds. The calculation uses NSLS-II operating conditions of 500~mA uniformly distributed over 1200 bunches, corresponding to 1.10~nC per bunch, with a vertical emittance of 30~pm$\cdot$rad and a coupling ratio of 4.2\%. Figure~\ref{Touschek_lifetime} shows the resulting lifetime distribution. The average calculated Touschek lifetime is approximately 12.4~h, which is adequate for routine operation under these assumptions.

\subsection{Lattice correction capability}

The replacement of two electromagnetic dipoles with permanent-magnet CBs reduces the local optics tunability. Nevertheless, the error study shows that the modified lattice can be corrected with the existing NSLS-II correction scheme. The orbit, beta functions, tunes, dispersion, and transverse coupling are restored to levels comparable to those of the nominal 17-ID lattice. The corrected lattice also achieves a sufficient DA and Touschek lifetime for off-axis injection and routine operation.

For additional flexibility required for beam-based alignment, optics correction, and lattice tuning studies, compact multipole correctors are planned to be installed upstream and downstream of each CB assembly. These corrector magnets are designed to compensate for residual magnetic errors of the CB assembly, including field strength deviations, harmonic errors, and alignment imperfections.

\begin{figure}[!htbp]
   \centering
   \includegraphics[width=0.7\columnwidth]{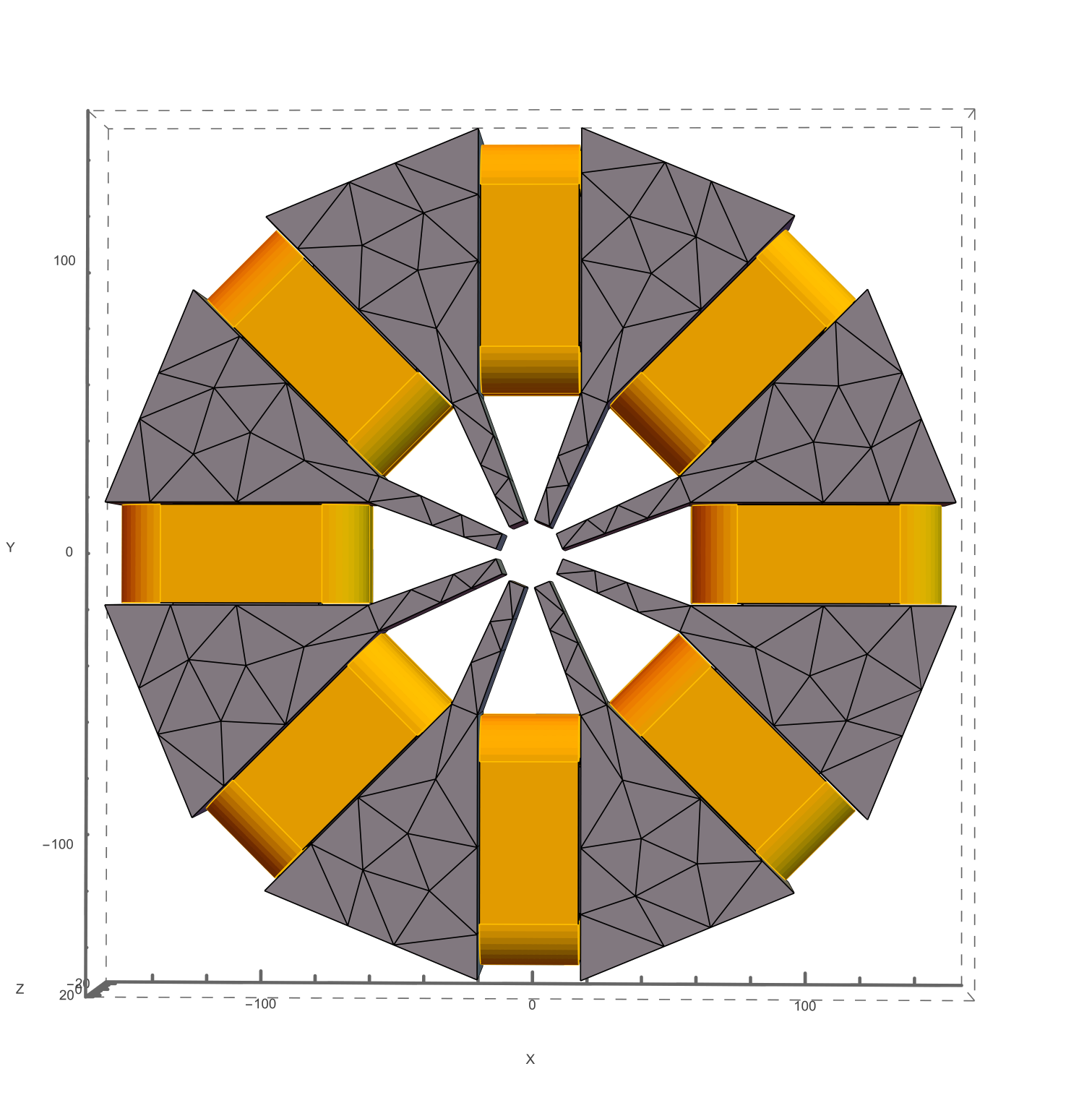}
   \caption{Conceptual design of the compact eight-pole corrector for the CB installation. Eight independently powered coils provide normal and skew multipole correction from dipole through octupole order within a longitudinal length of less than 50~mm.}
   \label{8-poles-corrector}
\end{figure}

The limited longitudinal space available in the lattice constrains the corrector magnet length to less than 50~mm. To satisfy this requirement, the proposed corrector, shown in Fig.~\ref{8-poles-corrector}, employs an eight-pole iron core with eight independently powered coils. This configuration enables the generation of both normal and skew multipole fields from the dipole through the octupole order. By independently controlling the current in each coil using a dedicated current combination strategy, the magnet can generate arbitrary combinations of normal and skew multipole components with the required polarity and strength.

The proposed eight-pole geometry incorporates a radial coil arrangement between adjacent poles, eliminating conventional coil end turns. This design significantly reduces the required longitudinal length while making efficient use of the available space, resulting in a compact, versatile, and highly integrated multipole corrector suitable for the stringent spatial constraints of the NSLS-II lattice with CBs.

Harmonic sextupoles are not used in the linear correction scheme. However, they are available for online tuning during commissioning.

\subsection{Magnetic cross-talk}
\label{sec:cross_talk}

Magnetic cross-talk is the interaction between adjacent magnets that causes deviations in the magnetic field and gradient strength from the nominal values of each individual magnet. For the CB assembly shown in Fig.~\ref{complex_bend_array}, the cross-talk primarily affects the focusing PMQ (CB-BQF), resulting in a 2.2\% reduction in the gradient and a 1.8\% reduction in the field. The defocusing PMQ (CB-BQD) experiences a more moderate effect, with a 0.06\% increase in gradient and a 0.95\% decrease in field. The dipole magnets (CB-B) exhibit a maximum field reduction of 1\%.

These deviations exceed the magnetic tolerances specified in Table~\ref{tab_error_params} and therefore require compensation. Although increasing the spacing between adjacent magnets would reduce cross-talk, this approach is not feasible because of the limited space available in the CB assembly.

Instead, the magnetic design is optimized to compensate for the cross-talk effects, reducing the residual field and gradient deviations to less than 0.5\%. In the final assembly, dedicated field and gradient adjustment plates, originally reserved for correcting permanent magnet imperfection and mechanical assembly errors, will be used to perform the final tuning and ensure that all magnets satisfy the specified field and gradient tolerances.

The machine error study shows that the lattice is correctable with the existing NSLS-II correction scheme when the errors within the tolerances. A detailed cross-talk analysis and the final magnet optimization will be reported separately.

\section{Experimental validation of the CB beam aperture}
\label{sec:Experimental validation of the CB beam aperture}

As part of the CB installation project, a dedicated scraper was designed, fabricated, and installed to validate the beam aperture requirements used in the CB design. The magnet bore radius is a key design parameter because it limits the achievable field and gradient. The size of the bore must be larger than the vacuum chamber and must provide sufficient clearance for mechanical alignment.

\begin{figure}[!htbp]
   \centering
   \includegraphics[width=0.9\columnwidth]{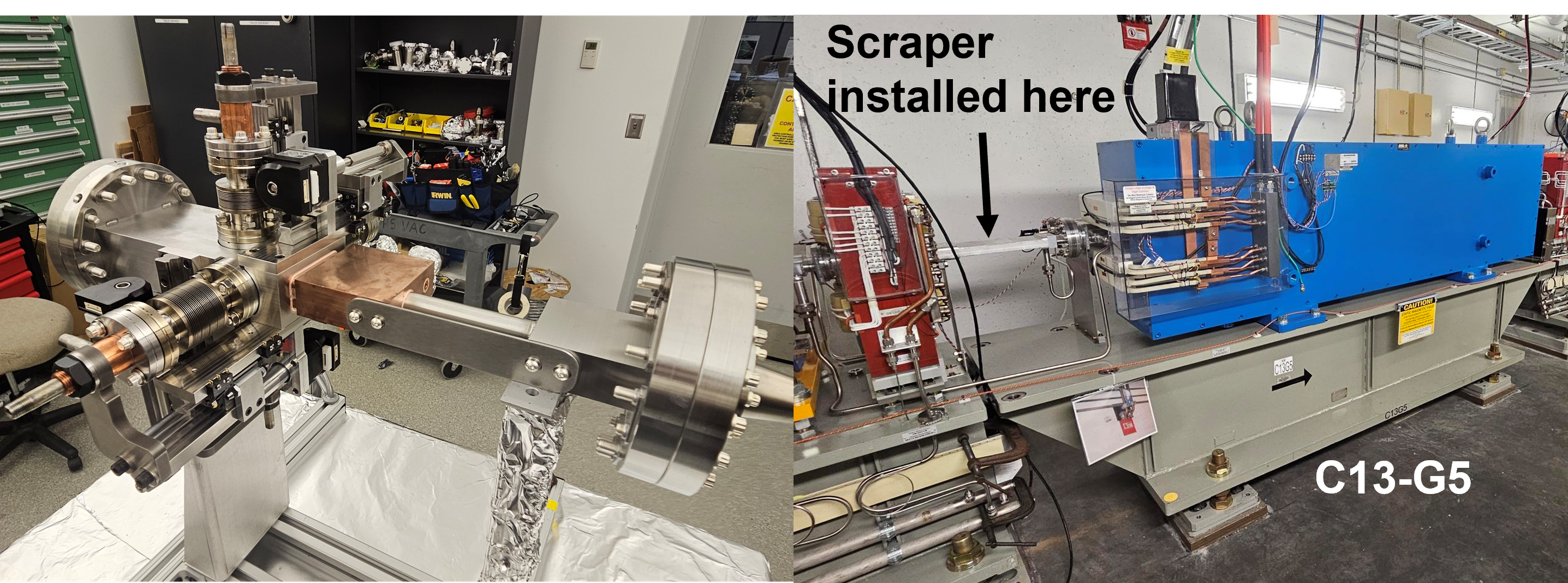}
   \caption{Dedicated scraper and its installation location on Girder 5 in Cell 13 adjacent to the planned CB installation location. The scraper allows independent adjustment of the horizontal and vertical beam apertures.}
   \label{CB_beam_scraper}
\end{figure}

As shown in Fig.~\ref{CB_beam_scraper}, the scraper was installed on Girder 5 in Cell 13, next to the planned CB location. It provides independent control of the horizontal and vertical apertures, allowing direct measurements of the aperture needed to preserve injection efficiency and beam lifetime.

\begin{figure}[!htbp]
   \centering
   \includegraphics[width=0.8\columnwidth]{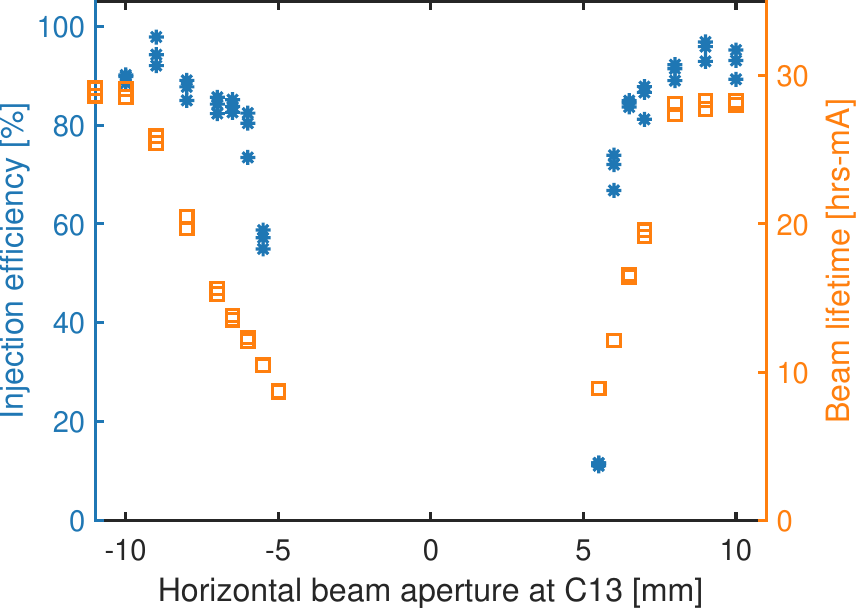}
   \includegraphics[width=0.8\columnwidth]{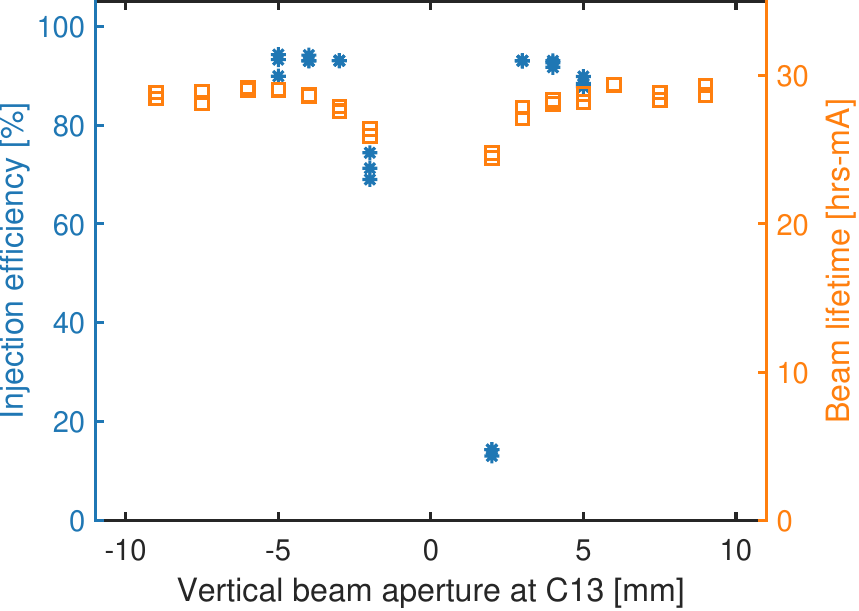}
   \caption{Measured injection efficiency and beam lifetime as functions of the horizontal (upper) and vertical (lower) beam apertures at Cell 13 (C13). The minimum required beam aperture is determined from the first observable degradation in either beam lifetime or injection efficiency.}
   \label{beam_aperture_scan}
\end{figure}

The measurements were performed using the operational 17-ID lattice, which is identical to the lattice used in the simulations. The scraper was moved step by step from the positive and negative sides in both transverse planes. Injection efficiency and beam lifetime were measured at each aperture, and each measurement was repeated three times. The results are shown in Fig.~\ref{beam_aperture_scan}.

Figure~\ref{beam_aperture_scan} shows that the beam lifetime remains nearly constant until the full horizontal aperture is reduced to approximately 18~mm, below which it begins to decrease. The injection efficiency remains nearly unchanged until a smaller aperture is reached. The minimum required aperture is defined by the first observable degradation in either beam lifetime or injection efficiency. Based on this criterion, the required full horizontal aperture is approximately 18~mm, in good agreement with the simulated aperture requirement for the CB design.

In the vertical plane, the aperture is mainly limited by the ID stay-clear requirement. The stay-clear aperture $g$ (minimum required full aperture) anywhere in the NSLS-II ring is~\cite{NEXTIII_DMB}: 
\begin{equation}
    g = A\sqrt{\beta_y},
\end{equation}
where $A=2.84\times10^{-3}~\sqrt{\mathrm{m}}$ and $\beta_y$ is the vertical beta function. At the scraper location, $\beta_y\approx18$~m, giving a minimum full aperture of approximately 12~mm. The measured beam lifetime begins to decrease when the vertical aperture is reduced below about $\pm6$~mm, while the injection efficiency remains nearly unchanged until the aperture is reduced further. Applying the same criterion gives a minimum required full vertical aperture of approximately 12~mm, consistent with the stay-clear requirement.

These measurements validate the horizontal and vertical beam aperture requirements used in the CB design. The measured thresholds agree with the simulation and operational stay-clear requirements, confirming that the selected 11.5-mm bore radius provides sufficient beam clearance while allowing the required magnetic fields and gradients.

\section{Conclusion}
\label{sec:Conclusion}

A modified NSLS-II 17-ID lattice has been developed for the planned replacement of two existing electromagnetic dipoles with permanent-magnet complex bends (CBs). The modified lattice preserves the existing ring geometry while satisfying the CB magnet design constraints. After nonlinear optimization, the lattice provides sufficient dynamic and momentum apertures for off-axis injection and routine operation.

The lattice performance was further evaluated using realistic machine imperfections based on the existing NSLS-II error model. The results show that the modified lattice can be corrected using the existing NSLS-II orbit, optics, and coupling correction scheme. After correction, the lattice achieves beam dynamics performance comparable to that of the nominal 17-ID lattice, with sufficient dynamic aperture, momentum acceptance, and Touschek lifetime for routine operation.

A dedicated scraper measurement was performed near the planned CB installation location to validate the beam aperture requirements. The measured horizontal and vertical aperture thresholds agree well with the design requirements and validate the selected CB bore radius.

This work presents the lattice design and beam dynamics studies for the planned CB installation at NSLS-II. The planned installation will provide the first experimental demonstration of the CB concept in an operating storage ring and provide valuable experience for the future development of a CB-based NSLS-II upgrade.

\begin{acknowledgments}
We would like to thank James Safranek (SLAC, USA) for his valuable suggestions and helpful discussions regarding this work.
This work is supported by the U.S. Department of Energy, Office of Basic Energy Sciences, under Contract No.~DE-SC0012704 and Field Work Proposal No. PS043.
\end{acknowledgments}

\section*{Data availability}
The data that support the findings of this study  are available upon reasonable request and subject to standard U.S. national laboratory data-sharing policies.

\bibliography{ref.bib}

\end{document}